\documentclass[10pt,leqno]{article}
\usepackage{graphicx} 

\usepackage{cite}
\usepackage{amsmath,amssymb,amsfonts}
\usepackage{algorithmic}
\usepackage{graphicx}
\usepackage{textcomp}
\usepackage{listings}
\usepackage[inkscapeformat=png]{svg}
\usepackage{balance}
\usepackage{authblk}
\usepackage{tabularx}

\usepackage[utf8]{inputenc}
\usepackage{csquotes}

\usepackage[T1]{fontenc}
\usepackage{geometry}
\usepackage{tikz}
\usepackage{circuitikz}

\usepackage{scalerel}

\usetikzlibrary{svg.path}

\definecolor{orcidlogocol}{HTML}{A6CE39}
\tikzset{
  orcidlogo/.pic={
    \fill[orcidlogocol] svg{M256,128c0,70.7-57.3,128-128,128C57.3,256,0,198.7,0,128C0,57.3,57.3,0,128,0C198.7,0,256,57.3,256,128z};
    \fill[white] svg{M86.3,186.2H70.9V79.1h15.4v48.4V186.2z}
                 svg{M108.9,79.1h41.6c39.6,0,57,28.3,57,53.6c0,27.5-21.5,53.6-56.8,53.6h-41.8V79.1z M124.3,172.4h24.5c34.9,0,42.9-26.5,42.9-39.7c0-21.5-13.7-39.7-43.7-39.7h-23.7V172.4z}
                 svg{M88.7,56.8c0,5.5-4.5,10.1-10.1,10.1c-5.6,0-10.1-4.6-10.1-10.1c0-5.6,4.5-10.1,10.1-10.1C84.2,46.7,88.7,51.3,88.7,56.8z};
  }
}

\newcommand\orcidicon[1]{\href{https://orcid.org/#1}{\mbox{\scalerel*{
\begin{tikzpicture}[yscale=-1,transform shape]
\pic{orcidlogo};
\end{tikzpicture}
}{|}}}}

\usepackage[hidelinks]{hyperref}

\tikzstyle{new style 0}=[fill=white, draw=black, shape=rectangle]
\tikzstyle{circle}=[fill=white, draw=black, shape=circle]

\tikzstyle{arrow}=[->]
\tikzstyle{bidirectional arrow}=[<->]
\tikzstyle{dashed line}=[-, dashed]
\tikzstyle{dashed bidirectional arrow}=[densely dashed, <->]
\tikzstyle{arrow dashed}=[dashed, ->]
\tikzstyle{red}=[-, thick, fill=none, draw={rgb,255: red,191; green,0; blue,64}]
\tikzstyle{dotted line}=[-, thick, dotted]
\tikzstyle{grey}=[-, draw={rgb,255: red,128; green,128; blue,128}]
\tikzstyle{filled}=[-, fill={rgb,255: red,44; green,174; blue,255}, opacity=0.2, draw={rgb,255: red,44; green,174; blue,255}]
\tikzstyle{green line}=[-, draw={rgb,255: red,28; green,191; blue,53}, thick]

\pgfdeclarelayer{nodelayer}
\pgfdeclarelayer{edgelayer}
\pgfsetlayers{edgelayer,nodelayer,main}

\title{Q-AIM: A Unified Portable Workflow for Seamless \\ Integration of Quantum Resources}

\author[ ]{%
Zhaobin Zhu\textsuperscript{1,*}, 
Cedric Gaberle\textsuperscript{2,*}, 
\\
Sarah M. Neuwirth\textsuperscript{1,3} 
Thomas Lippert\textsuperscript{2,3}, 
and Manpreet S. Jattana\textsuperscript{2}%

}

\affil[1]{Institute of Computer Science and Zentrum f\"ur Datenverarbeitung, Johannes Gutenberg University Mainz, D-55099 Mainz, Germany}
\affil[2]{Modular Supercomputing and Quantum Computing, Institute of Computer Science, Goethe University Frankfurt, D-60325 Frankfurt, Germany}
\affil[3]{Jülich Supercomputing Centre, Forschungszentrum Jülich GmbH, D-52428, Jülich, Germany}

\date{}

\begin{document}
\maketitle

\def\thefootnote{*}\footnotetext{These authors contributed equally to this work}

\begin{abstract}
Quantum computing (QC) holds the potential to solve classically intractable problems. Although there has been significant progress towards the availability of quantum hardware, a software infrastructure to integrate them is still missing. We present Q-AIM (Quantum Access Infrastructure Management) to fill this gap. Q-AIM is a software framework unifying the access and management for quantum hardware in a vendor-independent and open-source fashion.

Utilizing a dockerized micro-service architecture, we show Q-AIM's lightweight, portable, and customizable nature, capable of running on different hosting paradigms ranging from small personal computing devices to cloud servers and dedicated server infrastructure. Q-AIM exposes a single entry point into the host's infrastructure, providing secure and easy interaction with quantum computers on different levels of abstraction.  With a minimal memory footprint, the container is optimized for deployment on even the smallest server instances, reducing costs and instantiation overhead while ensuring seamless scalability to accommodate increasing demands. Q-AIM intends to equip research groups and facilities purchasing and hosting their own quantum hardware with a tool simplifying the process from procurement to operation and removing non-research related technical redundancies.
\end{abstract}


\section{Introduction}
\label{sec:introduction}
Quantum computing (QC), currently in its developmental phase, promises substantial acceleration of classical computations across various fields ranging from cryptography to materials science\cite{quantum_cryptography, quantum_cryptography_advancements, qc_materials_science, qc_applications}. Quantum computing scientists are constantly striving to overcome the limitations imposed by the current noisy intermediate-scale quantum (NISQ) era to fully realize quantum computing's potential. 

However, while large private-sector enterprises are advancing the field through their own hardware, software, and algorithmic developments, smaller academic research groups lack direct on-site access to such resources. Although corporations such as IBM \cite{ibm_pricing}, Google \cite{google_pricing}, and Amazon \cite{amazon_pricing} offer access to their own or hosted third-party infrastructure on a pay-to-use basis over the cloud, fundamental research is limited by restricted privilege policy and physical inaccessibility. Consequently, the acquisition of small-scale devices emerges as a viable solution to delve deeper into hardware and software enhancement studies, especially since the devices are getting cheaper. Yet, a critical challenge remains: the lack of a portable, open source, and easily integrable software solution for small-scale hardware integration and provision. 

Ultimately, procuring quantum hardware serves not only to enable deeper interaction with the device but also to facilitate its utilization on an abstract software level. This requires granting access to the resource over the host's network infrastructure, and possibly even beyond that, by a service either hosted in the  cloud or also on-premise, dependent on the requirements and capabilities. For instance, a device could be made accessible to external users, such as students, for educational purposes or to demonstrate advancements to a broader audience. Yet, the absence of a common, open-source integration platform forces researchers to spend valuable time and expertise developing such a solution on their own. Such efforts can detract from their primary focus of advancing scientific knowledge. A flexible, streamlined, and universally adaptable integration software is therefore crucial, not only to eliminate redundancies but also to ensure compatibility with existing workflows.

This work presents Q-AIM, a flexible, streamlined, and universally adaptable quantum integration workflow designed to address key challenges in quantum resource utilization, particularly for small enterprise and academic research groups. Typically, quantum systems are equipped with peripheral classical hardware providing a hardware- and vendor-dependent interface to the quantum computer, facilitating their use on an abstract software level. But, without a standardized, open-source platform, researchers face significant hurdles in integrating quantum systems into existing workflows. To eliminate redundancies and enhance compatibility of the necessary integration software solution, we make the following contributions: 

\begin{itemize}
    \item \textit{\textbf{Unified and Portable Platform}}: A Docker-based, microservice architecture ensures seamless deployment and scalability across various infrastructures e.g, on a local machine, server, and cloud \cite{docker}.
    \item \textit{\textbf{Flexible Access and Control}}: Offers resource access via multiple abstraction levels (from algorithmic to pulse-level) with a role-based permission scheme for secure and tailored utilization among diverse user groups.
    \item \textit{\textbf{Classical Workflow Integration}}: Standardized and flexible APIs enable easy hybrid computing, reproducibility, and cross-institution collaboration without major infrastructure changes.
    \item \textit{\textbf{Prototype Validation}}: A lightweight prototype demonstrated adaptability and efficient resource usage across on-premise  and cloud infrastructures, supporting broad research and educational application possibilities.
\end{itemize}

\section{Background and Related Work}
\label{sec::aorw}
Quantum computing offers great potential, but the integration of quantum hardware into classical workflows faces major challenges due to proprietary systems and lack of standardization. This chapter provides a brief overview of quantum instruction workflows and existing integration solutions.

\subsection{Quantum Instruction Workflow}
The instruction of any quantum hardware involves multiple steps, based on the level of abstraction, to match the desired logic to the operations exposed by the architecture. A stepwise workflow example from high-level algorithm definition to device-specific machine instructions is shown in Fig.~\ref{fig:qc_instruction_abstraction}, which includes the abstraction levels commonly used in quantum computing. 

The first step in every quantum execution is the definition of the algorithm as a quantum circuit (\textit{Circuit Definition}) making use of any arbitrary quantum operations in an abstract high-level software framework. Commonly used ones include Qiskit \cite{qiskit}, Cirq \cite{cirq}, and Braket \cite{braket_quantumAPI}, provided by IBM, Google, and Amazon respectively. This is the same principle as writing algorithms and using resources in classical computing, i.e. abstract away all hardware constraints and focus solely on the implementation of the desired logic. 

The closest quantum computing equivalent to the compilation is \textit{transpilation}, with an intermediate representation (IR) comparable to the executable \cite{IR-Review}. The transpilation introduces some form of optimization (\textit{HW-Independent IR}) and most importantly takes hardware constraints into account to fit the abstract level implementation to a backend (\textit{HW-Dependent IR})\cite{distributed_quantum_computing}. The latter is due to the native gate-set exposed by the quantum processor architecture, i.e. the logic of the high-level implementation must be represented using only a limited set of universal gates natively supported by the hardware in question, usually a few single-qubit rotation gates and a single two-qubit gate. BQSKit (Berkeley Quantum Synthesis Toolkit) \cite{bqskit} is an exemplary compiler framework to perform optimization and gate-set transpilation combining state-of-the-art algorithms in a stand-alone, end-to-end solution to reduce the depth of the quantum circuit. A more extensive comparison of quantum software development kits and compilers can be found in Ref. \cite{quantum_compiler_comparison}. Like the binary in classical computing, the IR can be of different formats. One of the most commonly used formats is OpenQASM \cite{openqasm2, openqasm3}, a quasi-assembly language for quantum computing. 

As in classical computer science, assembler is not yet an instruction at machine level, but is used to communicate with remote resources if used. Therefore, the last step of instruction modification is the translation to machine code (\textit{Machine Instructions}). In quantum computing, this oftentimes means microwave pulse modification where specific pulses modify the state of the quantum system likewise to an instruction in the high-level abstraction implementation.

\subsection{Accessing Quantum Resources}
\begin{figure}[t]
    \centering
    \includegraphics[width=0.7\columnwidth]{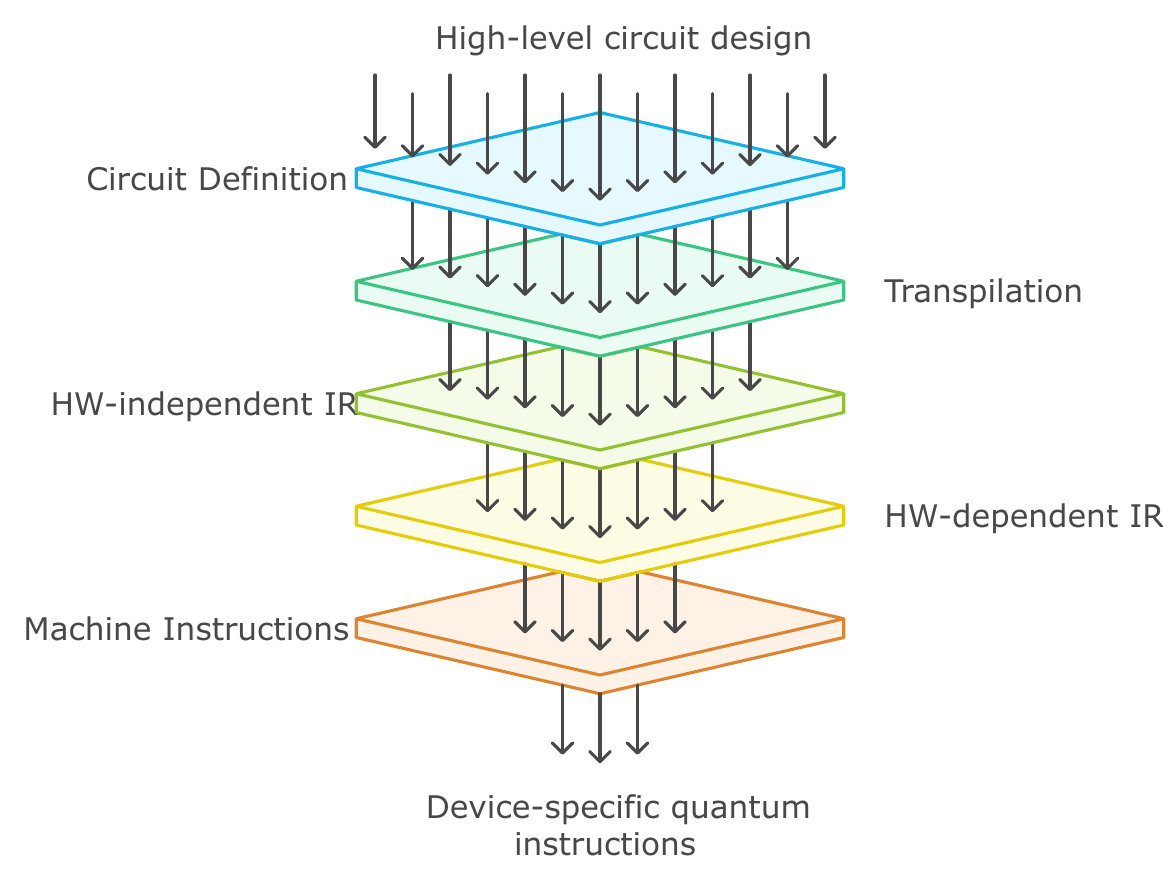}
   \caption{Instruction abstraction levels in quantum computing. From high-level circuit design (highest abstraction) to hardware-independent and hardware-specific intermediate representations (IR), ultimately becoming device-specific machine instructions (no abstraction) to be used with the quantum device.}
    \label{fig:qc_instruction_abstraction}
\end{figure}

For users to be able to execute algorithms on real quantum hardware, two parts must be taken into consideration. As depicted in Fig. \ref{fig:qc_instruction_abstraction} and discussed in the previous section, the first involves the transpilation of any algorithm or circuit to the underlying hardware in use. Usually, algorithms are developed and translated into quantum formalisms in a hardware-agnostic fashion. The high-level code abstraction must be transformed into a device-specific instruction set. 

The second part addresses with the access of the resources. As mentioned before, quantum computing providers often grant access to their devices over the cloud using an API interface. Therefore, accessing quantum computers as a form of highly specialized compute resource does not differ from accessing any other remote compute resource. Providers will most likely impose restrictions in terms of supported IR formats and the ways to interact. Invariably, they implement stringent security measures to safeguard their resources, mandating users to authenticate themselves via provided credentials and implement different levels of permissions. Typically, access to the protected resource is facilitated through an API, overseeing the bidirectional flow of data to and from the resource. In the workflow diagram depicted in Fig. \ref{fig:qc_instruction_abstraction}, the API call can happen at any level between the circuit definition and the machine instructions, depending on the services provided by bespoken API. The data flow from the resource to the user happens after execution, containing the result of the computation in a pre-defined format dependent on the provider.

\subsection{Analysis of Related Work}
Recent efforts aim to integrate quantum devices into classical computational resources, either by allowing them to be accessed during computation in a manner similar to GPUs (indirect access) \cite{accelerating_hpc_with_qc, quantum_acceleration, qc_for_hpc} or through an API service (direct access) \cite{quantum_cloud_reference_architecture, quantum_cloud_review, quantum_cloud_review2, qc_for_hpc, provider_study_and_api, quantumcomputingqiskit, braket_quantumAPI, cirq}. Ruefenacht et al.~\cite{quantum_acceleration} provide a fundamental overview of different integration architectures, from loosely- (on-premise) to tightly-coupled (on-chip), for quantum processors as accelerators for specific high-performance computing (HPC) workloads. Schulz et al.~\cite{accelerating_hpc_with_qc} particularly emphasize the necessity of developing an unified software stack to harness the strength of both radically different systems for the latter time and space shared integration scheme. However, Humble et al.~\cite{qc_for_hpc} state in their analysis of the different possibilities of integrating QC and HPC that \enquote{existing QC prototypes are based on loosely integrated client-server interactions that lack the sophistication or technological maturity to be used as accelerators}. But exactly these early prototypes are of particular interest for fundamental research done by academic groups and the center of possible application of the software proposed in this paper alike. Therefore, the focus for now is on the loose integration of (less sophisticated) quantum hardware. Such a solution in a corporation independent and open source fashion is the missing key to enable small scale device utilization and integration for academic purposes. This paper addresses the development of such a solution: a software platform that enables secure, low-overhead, and flexible access to quantum devices on different environments, such as on-premise and/or cloud infrastructures. Delivered as a Docker container, it ensures ease of deployment, hardware independence, and extensibility, creating a foundation for efficient and portable quantum computing research. 

Different research studies~\cite{qc_service_oriented, qfaas, serverless_cloud_qc} focusing on providing quantum computing as a service use dockerization for the same purposes. The services provided are usually some quantum application/algorithm, which abstract away the complexities of the hardware, simplify development and deployment, and allow for portability across different platforms. Grossi et al.~\cite{serverless_cloud_qc} proposes a decoupled function-as-a-service (FaaS) quantum computing integration, exposing a single HTTP API endpoint to provide an arbitrary quantum service, which is offloaded to the IBM backend, thereby separating provision and usage. Nguyen et al.~\cite{qfaas} expand on the idea, pointed out the vendor lock-in effect and extended the capabilities of the serverless model to different backend platforms. Overall, the goal is a higher abstraction for ease of use in enterprise scenarios. This work addresses the exact opposite. We provide services also to be hardware-agnostic, portable, and flexible, but with the main purpose of allowing fine-grained access to the device on levels of less abstraction, e.g. interaction on pulse-level. Researchers are dependent on a low-level access since the focus is not on the usage through quantum algorithms as a service but on being able to accurately manipulate certain workflows and interactions in the pursuit of knowledge extraction. 
 
During the time of writing, a paper~\cite{BECK2024} emphasizing the necessity and challenges of a hardware-agnostic framework integrating hardware, software, workflows, and user interfaces, also with particular focus on academia to foster a synergistic environment for quantum and classical computing research was published. The main difference to the work in this paper is that they present developments in software for a hybrid quantum computing approach integrating quantum computers as accelerators to complement high performance computing systems. These advancements also primarily target large academic institutions already running sophisticated HPC systems as a form to further improve computations of the specific field of interest. Our work focuses on enabling small (research) groups to integrate, directly access, and make own quantum devices publicly accessible in an unified fashion, streamlining the process after procurement until usage, e.g. investigative research.

\section{Design Consideration}
\label{sec:hlc}

\begin{figure}[t]
    \centering
    \resizebox{0,8\textwidth}{!}{
        \input{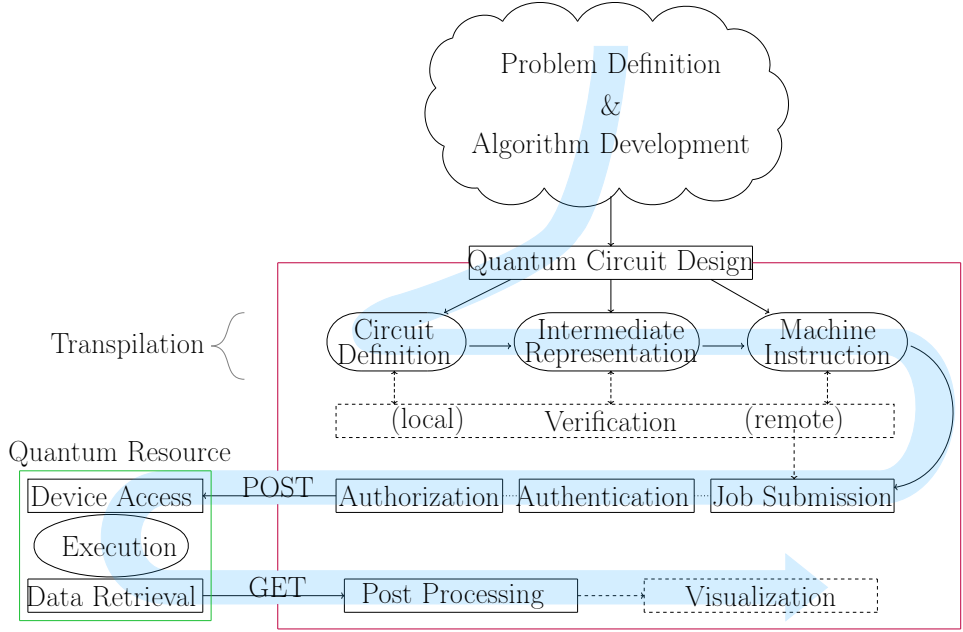}
    }
    \caption{Overview of the quantum computation workflow. The larger, red box (right) indicates classical hardware, the smaller, green one (left) the quantum system. The problem is defined as quantum circuit at any abstraction level and undergoes transpilation until machine instruction level is reached. Communication between the classical and quantum system is facilitated through API calls.}
    \label{fig:quantum_workflow}
\end{figure}

As described in Section~\ref{sec::aorw}, the integration of quantum computing resources into existing research and industrial workflows requires careful orchestration of software across multiple domains. However, today's quantum computing solutions are often associated with specialized hardware and proprietary environments, limiting their applicability to on-premise installations and creating major barriers to entry. This is a significant obstacle, especially for smaller institutions and educational facilities.  

To address these challenges, we propose Q-AIM, a standardized, portable workflow and a corresponding software implementation that enables seamless integration of quantum resources into existing infrastructures. As shown in Fig.~\ref{fig:quantum_workflow}, the required process steps are highlighted in blue, the involved services on the classical hardware side by the red box and the vendor-dependent quantum resources by the green box. 

With Q-AIM, quantum circuit design can be performed at various levels of abstraction. We also introduce an optional simulation step that serves as a verification phase before running computations on potentially busy remote quantum computers. All services indicated in the red box are abstractions that can be made available within the proposed software. Rather than enforcing a single standardized software framework, we allow each component or service to be replaced or customized, thereby giving both users and hardware providers granular control over implementation details. This flexibility accommodates diverse requirements and preferences. Because there are multiple execution environments and a wide range of components or services, each with distinct access control needs, we rely on well-established Representational State Transfer (REST) APIs, using common calls such as \textit{POST} and \textit{GET} to ensure broad compatibility across different software languages, platforms, and hardware architectures.

Another key design consideration is the decoupling of quantum hardware, ensuring that the software layer operates independently of specific hardware, thereby minimizing manufacturer dependency. Consequently, tasks such as qubit manipulation, machine instruction execution, and qubit state measurement are handled directly by the quantum hardware and its peripherals. Together with the microservices, these components form the backend for Q-AIM.

\section{Methodology}
The key principle of the proposed approach is to ensure that classical workflows remain largely unaffected by the introduction of quantum computers. Instead of having to rebuild or heavily modify pre-existing computational frameworks, end-users can embed quantum tasks and pipelines into their established processes. With this design approach, a quantum resource can be quickly adapted with minimal changes, while maximizing the advantages of quantum computing. The effectiveness and versatility of the proposed system are underpinned by four core methodologies:

\begin{itemize}
    \item Fully Integrated Classical Workflow
    \item Encapsulated System Architecture
    \item Micro-Service-Based Software Architecture
    \item Flexible and Fine-Grained User Management
\end{itemize}

\subsection{Fully Integrated Classical Workflow}
Based on the design consideration as mentioned in Section~\ref{sec:hlc}, the classical quantum computing workflow, i.e. the steps from algorithm definition to machine instruction, is a multi-stage process that spans tasks from algorithm development to analysis and optimization. Depending on the manufacturer and application scenarios, the code often needs to be compiled into an appropriate representation, such as gate-level or pulse-level instructions, to execute on a quantum computer. To allow users to operate at different levels of abstraction, it is crucial to account for these variations during the integration workflow.

To support this flexibility and maintain vendor independence, the entire classical quantum computing workflow is treated as a black box and integrated as a unified entity within our infrastructure. This abstraction ensures seamless interaction between the classical and quantum workflows without requiring users to manage low-level specifics or adapt to API changes, thereby enhancing usability and interoperability. Therefore, the classical workflow is incorporated into our integration pipeline as a self-contained component and augmented with additional functionality. These functionalities range from custom user management, authentication services, and access control to result visualization and system monitoring. This approach allows users to work with different programming languages at different levels of abstraction while taking advantage of the unique features of different quantum hardware backends. It also supports adaptability to emerging quantum computing platforms, ensuring that the architecture is future-proof. 

\subsection{Encapsulated System Architecture}
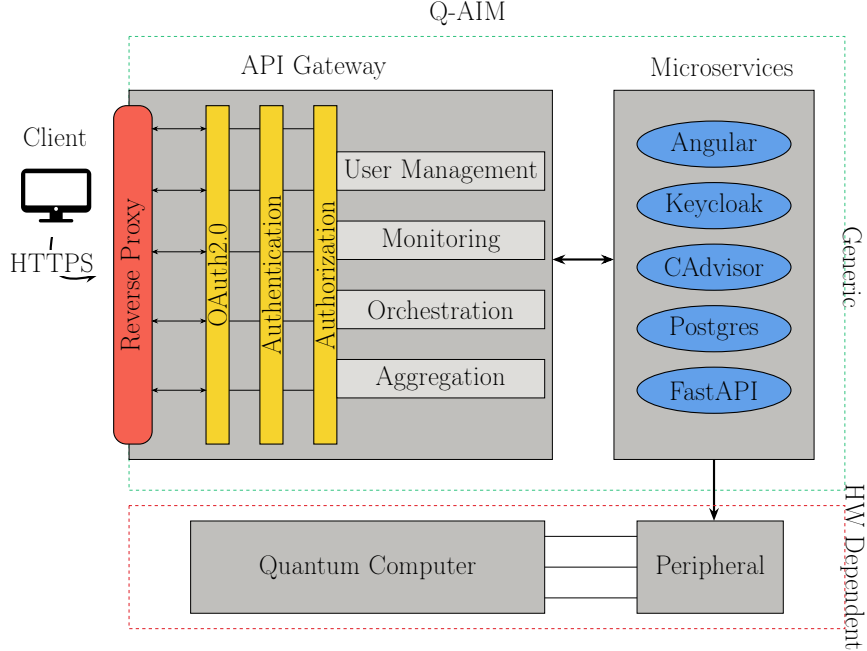
\begin{figure}[tbp]
\centering
\resizebox{0.7\columnwidth}{!}{
\begin{circuitikz}
\tikzstyle{every node}=[font=\Huge]
\draw [ color={rgb,255:red,46; green,194; blue,126} , dashed] (0.25,16.75) rectangle  (23.5,2);
\draw [ fill={rgb,255:red,192; green,191; blue,188} ] (16,15) rectangle (22.5,3);
\draw [ fill={rgb,255:red,192; green,191; blue,188} ] (0.25,15) rectangle (14,3);
\draw [ fill={rgb,255:red,98; green,160; blue,234} , line width=1pt ] (19.25,13.25) ellipse (2.5cm and 0.75cm) node {\Huge Angular} ;
\draw [ fill={rgb,255:red,98; green,160; blue,234} , line width=1pt ] (19.25,11.25) ellipse (2.5cm and 0.75cm) node {\Huge Keycloak} ;
\draw [ fill={rgb,255:red,98; green,160; blue,234} , line width=1pt ] (19.25,9.25) ellipse (2.5cm and 0.75cm) node {\Huge CAdvisor} ;
\draw [ fill={rgb,255:red,98; green,160; blue,234} , line width=1pt ] (19.25,7.25) ellipse (2.5cm and 0.75cm) node {\Huge Postgres} ;
\draw [ fill={rgb,255:red,246; green,97; blue,81} , line width=1pt , rounded corners = 12.0, ] (-0.25,14.5) rectangle  node {\Huge \rotatebox{90}{Reverse Proxy}} (1,3.5);
\draw [ fill={rgb,255:red,98; green,160; blue,234} , line width=1pt ] (19.25,5.25) ellipse (2.5cm and 0.75cm) node {\Huge FastAPI} ;
\draw [line width=2pt, <->, >=Stealth] (14,9.5) -- (16,9.5);
\draw [ fill={rgb,255:red,246; green,211; blue,45} , line width=1pt ] (2.75,14.5) rectangle  node {\Huge \rotatebox{90}{OAuth2.0}} (3.5,3.5);
\draw [ fill={rgb,255:red,246; green,211; blue,45} , line width=1pt ] (4.5,14.5) rectangle  node {\Huge \rotatebox{90}{Authentication}} (5.25,3.5);
\draw [ fill={rgb,255:red,246; green,211; blue,45} , line width=1pt ] (6.25,14.5) rectangle  node {\Huge \rotatebox{90}{Authorization}} (7,3.5);
\draw [short] (3.5,13.75) -- (4.5,13.75);
\draw [short] (3.5,11.75) -- (4.5,11.75);
\draw [short] (3.5,9.75) -- (4.5,9.75);
\draw [short] (3.5,7.5) -- (4.5,7.5);
\draw [short] (3.5,5.25) -- (4.5,5.25);
\draw [short] (5.25,13.75) -- (6.25,13.75);
\draw [short] (5.25,11.75) -- (6.25,11.75);
\draw [short] (5.25,9.75) -- (6.25,9.75);
\draw [short] (5.25,7.5) -- (6.25,7.5);
\draw [short] (5.25,5.25) -- (6.25,5.25);
\draw [ fill={rgb,255:red,222; green,221; blue,218} ] (7,13) rectangle  node {\Huge User Management} (13.75,11.75);
\draw [ fill={rgb,255:red,222; green,221; blue,218} ] (7,10.75) rectangle  node {\Huge Monitoring} (13.75,9.5);
\draw [ fill={rgb,255:red,222; green,221; blue,218} ] (7,8.5) rectangle  node {\Huge Orchestration} (13.75,7.25);
\draw [ fill={rgb,255:red,222; green,221; blue,218} ] (7,6.25) rectangle  node {\Huge Aggregation} (13.75,5);
\node [font=\Huge] at (6.25,15.75) {API Gateway};
\node [font=\Huge] at (19.5,15.75) {Microservices};
\node [font=\Huge] at (11.25,17.5) {Q-AIM};
\draw [line width=2pt, ->, >=Stealth] (-2.25,10.25) .. controls (-2.5,8.75) and (-1.75,8.75) .. (-0.75,9) node[pos=0.25, fill=white]{HTTPS};
\draw [<->, >=Stealth] (1,13.75) -- (2.75,13.75);
\draw [<->, >=Stealth] (1,11.75) -- (2.75,11.75);
\draw [<->, >=Stealth] (1,9.75) -- (2.75,9.75);
\draw [<->, >=Stealth] (1,7.5) -- (2.75,7.5);
\draw [<->, >=Stealth] (1,5.25) -- (2.75,5.25);
\node [font=\Huge] at (-2.15,13.5) {Client};
\node (tikzmaker) [shift={(1.375, -1.375)}] at (-3.5,13) {\includegraphics[width=2.75cm]{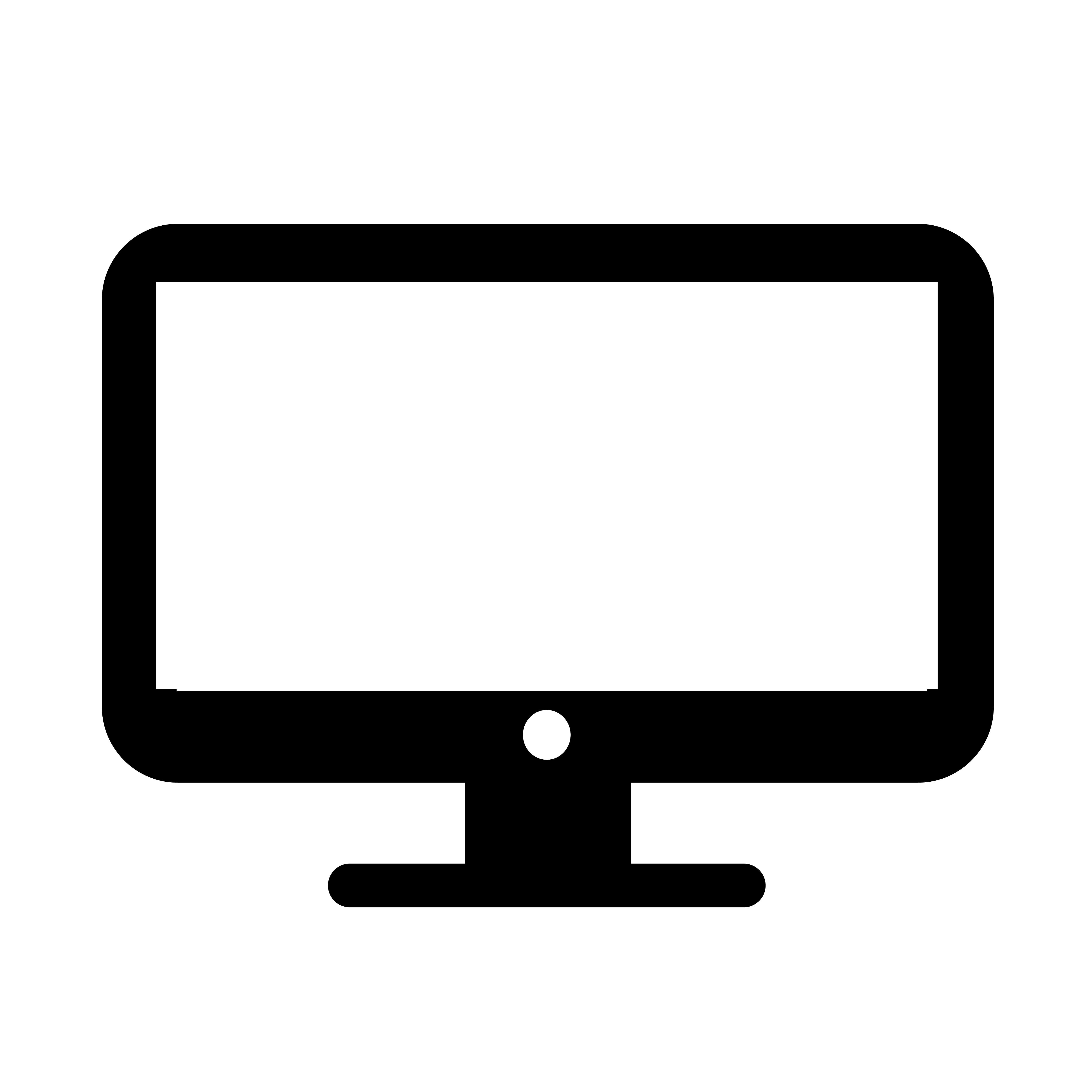}};
\draw [ fill={rgb,255:red,192; green,191; blue,188} ] (21.5,1) rectangle  node {\Huge Peripheral} (16.75,-2);
\draw [ fill={rgb,255:red,192; green,191; blue,188} ] (13.75,1) rectangle  node {\Huge Quantum Computer} (2.25,-2);
\draw [short] (13.75,0.5) -- (16.75,0.5);
\draw [short] (13.75,-1.5) -- (16.75,-1.5);
\draw [short] (13.75,-0.5) -- (16.75,-0.5);
\draw [line width=2pt, ->, >=Stealth] (19.25,3) -- (19.25,1);
\draw [ color={rgb,255:red,224; green,27; blue,36} , dashed] (0.25,1.5) rectangle  (23.5,-2.5);
\node [font=\Huge] at (23.75,9.25) {\rotatebox{270}{Generic}};
\node [font=\Huge] at (23.75,-0.5) {\rotatebox{270}{HW Dependent}};
\end{circuitikz}
}
\caption{Microservice-based architecture of Q-AIM. It facilitates secure client interactions via HTTPS and a reverse proxy, providing access to quantum systems through a structured microservice architecture. The API Gateway manages authentication, authorization, and orchestration, while the microservices provide the software's functionalities.}
\label{fig:system_architecture}
\end{figure}

To enable a standardized and transparent quantum computing workflow, we rely on an encapsulated system architecture that decouples the software layer from the underlying quantum computing hardware. This architecture acts as an abstraction layer that simplifies and hides the complexity of the individual components. 
As shown in Fig.~\ref{fig:system_architecture}, the system is divided into two key segments: \emph{the Q-AIM software} and \emph{the quantum computing hardware}. Given that the Q-AIM framework provides users with exclusive access to quantum resources via various microservices (see \ref{sub::mssa}), the central component of this system architecture is the API gateway, which serves as the primary entry point for users. It abstracts the underlying microservices and prevents direct access or communication between clients and service components. This isolation significantly simplifies implementation for both clients and microservice applications, as the complexity of the application is decoupled from its clients.
Another important element is the Reverse Proxy. The API gateway can be understood as a superset of the reverse proxy and offers additional functions that go beyond the simple forwarding of requests. The reverse proxy assigns the physical ports to those of the encapsulated environment and acts as an intermediary that communicates with the server on behalf of the client(s), forwards requests and returns responses. The proxy is located at the edge of the API gateway, which centralizes the processing of API requests and enforces additional security policies such as authentication, authorization and access control, as well as other functions not covered by the microservices.

As a result, the entire architecture is generic, portable, and easily extendable. It provides a standardized way for those to communicate, interact, and be managed within the environment, thus allows for modularity, scalability, and adaptability, making it possible to integrate the services seamlessly while maintaining a consistent and manageable architecture.

\subsection{Micro-Service-Based Software Architecture}
\label{sub::mssa}
To meet the challenge of a standardized, portable integration workflow, in this work we develop a microservice-based software architecture that enables quantum computing hardware to be integrated into existing and future infrastructures in a consistent manner. A key aspect for Q-AIM is therefore portability and transparency. 

To achieve this, the software requires the ability to run in a virtualized and isolated environment. Lightweight virtualization technologies, i.e. containers such as Docker~\cite{merkel2014docker} or Apptainer~\cite{kurtzer2021hpcng}, are highly portable, which means that they can be easily run on different operating systems and infrastructures. The isolated nature of container virtualization also ensures that all required dependencies are bundled in the container and services can be quickly deployed and replicated on different hosts. As container-based software deployment is typically based on a microservice architecture, the functionality of the software can be customized and extended according to user-specific requirements. This gives Q-AIM greater versatility and adaptability, which is beneficial for research institutions and companies alike.

Overall, Q-AIM's microservices-based architecture not only reduces the dependency on specific vendors, but also allows researchers and developers to transfer and scale their work to different environments \cite{moreau2023containers}. This is particularly important for reproducibility and enables the building of a community that promotes the exchange of ideas, best practices and resources to further advance the development of quantum computing technology.

\subsection{Flexible and Fine-Grained User Management}
Another key challenge is managing access from different environments with corresponding user affiliations. Users can generally be categorized into internal and external groups, each requiring specific levels of access to quantum resources. For example, a physicist conducting physical experiments on a quantum computer needs easy access to enter signals or waveforms. In contrast, users from business or other fields usually require high-level access to test their algorithms or circuits on the quantum computer.

To enable fine-grained access control to quantum resources and flexible user management, it is essential to integrate different user groups into a single infrastructure, manage them effectively and meet their different access requirements. This requires the integration of the industry standard LDAP~\cite{sermersheim2006rfc} protocol into our solution for authenticating internal users. In addition, the system should support the creation and management of a special user database for external users to ensure seamless integration and secure access for all user types. As interaction with quantum computing resources takes place exclusively via the API gateway, Q-AIM enables authentication for different user groups and supports fine-grained authorization, ensuring that users can only interact with the resources that correspond to their assigned roles.

\section{Prototype Implementation}
In the following, we present an early prototype implementation of our portable, unified, and generic quantum computing integration workflow. The integration of self-written or third-party libraries as a service in the example implementation of our microservice architecture underlines the aforementioned adaptability. Similarly, other entities can implement different services specific for their use cases.

\subsection{Container-based Deployment}
From the high-level system architecture shown in Fig.~\ref{fig:system_architecture}, it is clear that deploying the Q-AIM application requires a complex environment with a number of microservices working together. To improve transparency and portability in the deployment process, Docker containers are used to ensure consistency. Also, a Docker Compose file is used to simplify the management of multiple microservices and their dependencies within the application. Consequently, this approach facilitates the deployment of the entire application environment with a single command, i.e. \textit{docker compose up}.

\begin{lstlisting}[
 float=t,  
  basicstyle=\footnotesize\ttfamily,
  caption={Overview of the microservices and their images in the docker compose file.},
  captionpos=b,
  label=list:docker
]
                services:
                    database:
                        image: postgres
                        ...
                    authentification:
                        depends_on:
                            - database
                        image: jboss/keycloak:11.0.3
                        ...
                    Q-AIM-API:
                        image: fastapi:dev
                        ...
                    Q-AIM-Frontend:
                        image: Q-AIM:dev
                        ...
                    reverse-proxy:
                        image: nginx:alpine
                        ...
                    monitoring:
                        image: gcr.io/cadvisor/cadvisor:latest
                        ...
                volumes:
                    ...
\end{lstlisting}

To provide an overview of the main services of Q-AIM, as shown in Listing~\ref{list:docker}, the services are described below:

\begin{itemize}
    \item \textbf{Database Service}: This initiates a PostgreSQL database utilizing the official Postgres Docker image~\cite{PostgreSQLDocker}. To ensure persistent storage of the database data, a Docker volume is created alongside.
    
    \item \textbf{Authentication Service}: Utilizing the official Keycloak Docker image~\cite{KeycloakDocker}, this service delivers identity and access management functionalities. It relies on the database service and necessitates a Keycloak configuration file. For illustrative purposes, environment variables for the Keycloak administrator user, password and other settings are also configured via the docker compose file.
    
    \item \textbf{Q-AIM-API Service}: This employs the Docker image fastapi:dev and is built using a custom Dockerfile, which sets up an environment tailored for a FastAPI~\cite{fastapi} application and installs specific dependencies.
    
    \item \textbf{Q-AIM-Frontend Service}: Built upon the Q-AIM:dev Docker image using a custom Dockerfile,  this Dockerfile ensures that the actual Angular Web-Application ~\cite{Angular} is built in a Node.js~\cite{cantelon2014node} environment and then the resulting build is deployed within an NGINX ~\cite{reese2008nginx} container. The NGINX container is used to serve the static files of the Angular application and provide the configuration for the web server.
    
    \item \textbf{Reverse Proxy Service}: Based on the Docker image nginx:alpine, this service initializes an NGINX proxy server. Configured with a corresponding configuration file and SSL certificates, the proxy server forwards incoming requests to various services provided within Docker containers.
    
    \item \textbf{Monitoring Service (Optional)}: Leverages the official CAdvisor Docker image~\cite{cAdvisorDocker} to efficiently gather and present container statistics. To facilitate access to files or directories within the host system, it is imperative to include relevant directories or files from the host within the container. For instance, directories such as /sys/ need to be mounted for this service.
\end{itemize}

Overall, the division of microservices illustrates the basic principles of modern software development and architecture. This approach promotes customizability, scalability, security and reproducibility in application deployment. By using Docker and Docker Compose, both developers and professionals can seamlessly adapt Q-AIM to their specific requirements and deploy it efficiently in their infrastructure.

\subsection{Authentication Workflow}\label{sec:authentication_workflow}

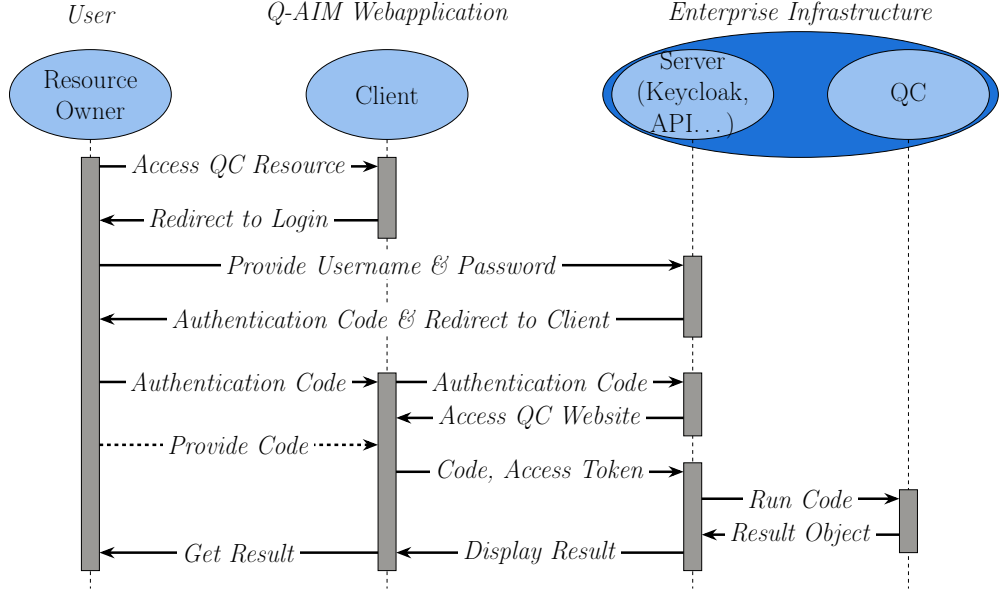
\begin{figure}[tbp]
\centering
\resizebox{0.8\textwidth}{!}{%
\begin{circuitikz}
\tikzstyle{every node}=[font=\huge, align=center]
\draw [ fill={rgb,255:red,28; green,113; blue,216} , line width=0.3pt ] (24.25,12.5) ellipse (5.5cm and 1.75cm);
\draw [ fill={rgb,255:red,153; green,193; blue,241} ] (4.5,12.5) ellipse (2.25cm and 1.25cm) node {Resource\\Owner} ;
\draw [ fill={rgb,255:red,153; green,193; blue,241} ] (12.75,12.5) ellipse (2.25cm and 1.25cm) node {Client} ;
\draw [ fill={rgb,255:red,153; green,193; blue,241} ] (21.25,12.5) ellipse (2.25cm and 1.25cm) node {Server\\ (Keycloak,\\API\dots)} ;
\draw [ fill={rgb,255:red,153; green,193; blue,241} ] (27.25,12.5) ellipse (2.25cm and 1.25cm) node {QC} ;
\draw [dashed] (4.5,11.25) -- (4.5,-1.25);
\draw [dashed] (12.75,11.25) -- (12.75,-1.25);
\draw [dashed] (21.25,11.25) -- (21.25,-1.25);
\draw [dashed] (27.25,11.25) -- (27.25,-1.25);
\draw [ fill={rgb,255:red,154; green,153; blue,150} , line width=0.2pt ] (4.25,10.75) rectangle (4.75,-0.75);
\draw [ fill={rgb,255:red,154; green,153; blue,150} , line width=0.2pt ] (12.5,10.75) rectangle (13,8.5);
\draw [line width=2pt, ->, >=Stealth] (4.75,10.5) -- (12.5,10.5)node[pos=0.5, fill=white]{\textit{Access QC Resource}};
\draw [line width=2pt, ->, >=Stealth] (12.5,9) -- (4.75,9)node[pos=0.5, fill=white]{\textit{Redirect to Login}};
\draw [ fill={rgb,255:red,154; green,153; blue,150} , line width=0.2pt ] (21,8) rectangle (21.5,5.75);
\draw [line width=2pt, ->, >=Stealth] (4.75,7.75) -- (21,7.75)node[pos=0.5, fill=white]{\textit{Provide Username \& Password}};
\draw [line width=2pt, ->, >=Stealth] (21,6.25) -- (4.75,6.25)node[pos=0.5, fill=white]{\textit{Authentication Code \& Redirect to Client}};
\draw [ fill={rgb,255:red,154; green,153; blue,150} , line width=0.2pt ] (12.5,4.75) rectangle (13,-0.75);
\draw [ fill={rgb,255:red,154; green,153; blue,150} , line width=0.2pt ] (21,4.75) rectangle (21.5,3);
\draw [ fill={rgb,255:red,154; green,153; blue,150} , line width=0.2pt ] (27,1.5) rectangle (27.5,-0.25);
\draw [line width=2pt, ->, >=Stealth] (4.75,4.5) -- (12.5,4.5)node[pos=0.5, fill=white]{\textit{Authentication Code}};
\draw [line width=2pt, ->, >=Stealth] (13,4.5) -- (21,4.5)node[pos=0.5, fill=white]{\textit{Authentication Code}};
\draw [line width=2pt, ->, >=Stealth] (21,3.5) -- (13,3.5)node[pos=0.5, fill=white]{\textit{Access QC Website}};
\draw [line width=2pt, ->, >=Stealth] (13,2) -- (21,2)node[pos=0.5, fill=white]{\textit{Code, Access Token}};
\draw [line width=2pt, ->, >=Stealth] (27,0.25) -- (21.5,0.25)node[pos=0.5, fill=white]{\textit{Result Object}};
\draw [line width=2pt, ->, >=Stealth] (12.5,-0.25) -- (4.75,-0.25)node[pos=0.5, fill=white]{\textit{Get Result}};
\draw [line width=2pt, ->, >=Stealth, dashed] (4.75,2.75) -- (12.5,2.75)node[pos=0.5, fill=white]{\textit{Provide Code}};
\draw [ fill={rgb,255:red,154; green,153; blue,150} , line width=0.2pt ] (21,2.25) rectangle (21.5,-0.75);
\draw [line width=2pt, ->, >=Stealth] (21.5,1.25) -- (27,1.25)node[pos=0.5, fill=white]{\textit{Run Code}};
\draw [line width=2pt, ->, >=Stealth] (21,-0.25) -- (13,-0.25)node[pos=0.5, fill=white]{\textit{Display Result}};
\node [font=\huge] at (24.25,14.75) {\textit{Enterprise Infrastructure}};
\node [font=\huge] at (12.75,14.75) {\textit{Q-AIM Webapplication}};
\node [font=\huge] at (4.5,14.75) {\textit{User}};
\end{circuitikz}
}%
\caption{Representation of the first authentication process for an approved user attempting to run code on a protected quantum computing resource.}
\label{fig:authentication}
\end{figure}

An exemplary workflow accessing a quantum device as protected resource is depicted in Fig.~\ref{fig:authentication}. During the user's initial access, they are required to provide their credentials. Only after the identity and access management tool Keycloak validates the provided credentials and returns an authentication code, including an access token holding information about the authenticated user's roles and permissions, an ID token with general information about the authenticated user, and a refresh token, does the user gain access to the quantum computer frontend component on the Angular webapplication. Provided quantum code of the user on the frontend component serves as input data to the API endpoint managing access to the protected quantum resource. The API therefore validates the provided authentication code at the identity and access management tool and checks the user's permissions in the access token. If the user is permitted, it controls the bidirectional flow to and from the quantum resource. Lastly, the result is displayed on the frontend web application.

\subsection{Q-AIM User Interface}
\begin{figure}[htbp]
\centerline{\includegraphics[width=\textwidth]{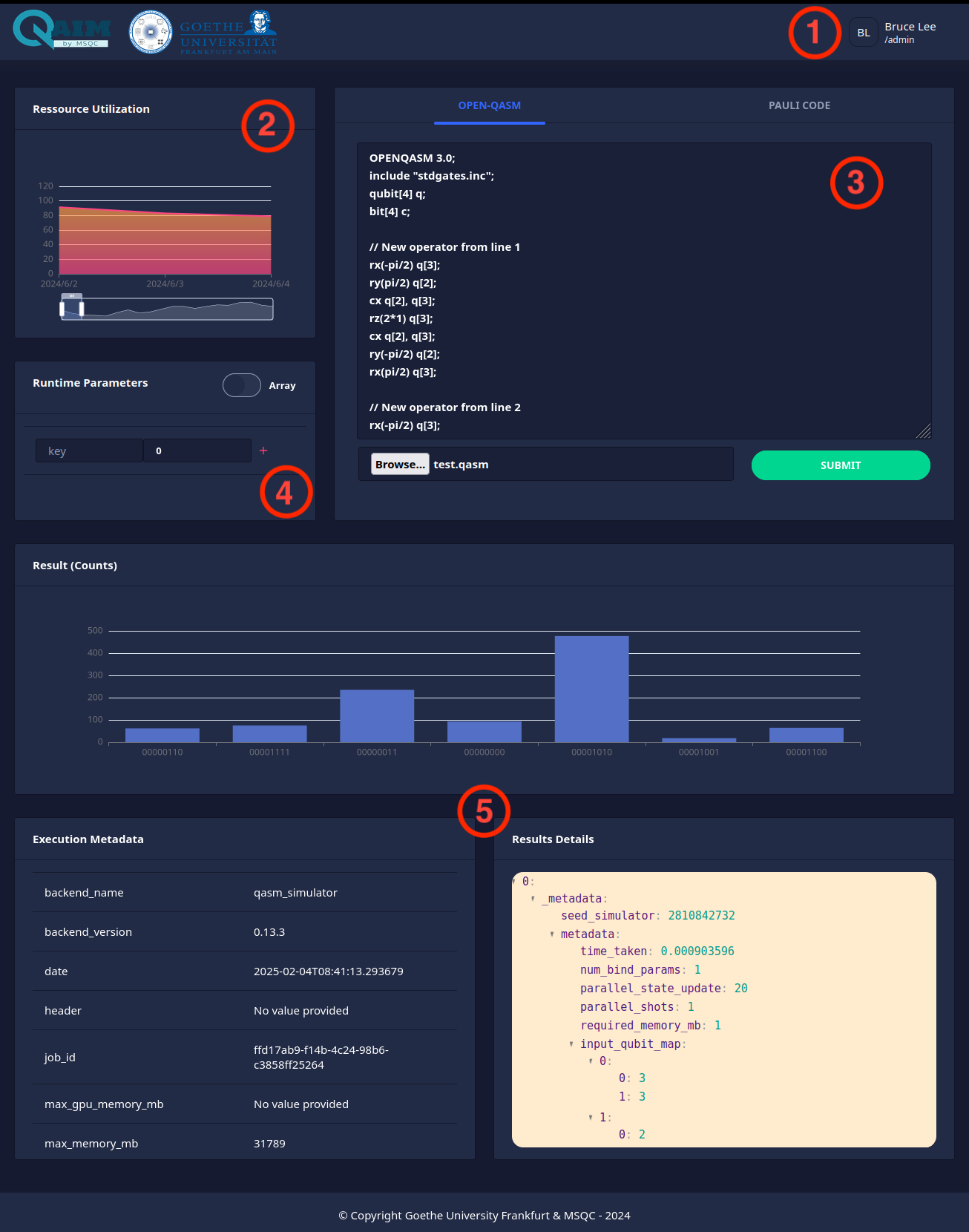}}
\caption{Q-AIM Web User Interface. Users can provide code and runtime parameters in different formats, monitor resource utilization, and visualize results and metadata.}
\label{fig::ui}
\end{figure}

The Q-AIM frontend serves as a user-friendly gateway to access quantum computing resources. To safeguard the underlying endpoints and enable fine-grained permissions management, integration of the Keycloak service and authentication functionality has been embedded within the Angular application. As can be seen from the Fig. \ref{fig::ui} \raisebox{.5pt}{\textcircled{\raisebox{-.9pt} {1}}}, users must be authenticated to access certain resources and have certain permissions. Furthermore, the authorization framework's distinction between groups and roles facilitates the assignment of users to various domains, institutions, and systems, allowing for the allocation of grouping-specific roles. To exemplify the granularity of rights management, the prototype establishes two groups, i.e., internal and external and each featuring user or admin roles. 
Depending on whether the user is already authenticated via the authentication server, the user is either redirected to the login page to process the authentication workflow as shown in Fig. \ref{fig:authentication} or to the interface for the corresponding compute resources, as shown in Fig.~\ref{fig::ui}. 

A standardized user interface ensures a seamless workflow for accessing different backend functionalities. As can be seen in \raisebox{.5pt}{\textcircled{\raisebox{-.9pt} {2}}}, the resource utilization of the respective quantum resource is displayed. \raisebox{.5pt}{\textcircled{\raisebox{-.9pt} {3}}} shows, Q-AIM currently supports OpenQASM source code or Pauli representation, a format introduced in \cite{cedric_thesis} and parsed into OpenQASM by a library made available as binding at \cite{qasm_parser}, as input. The Pauli representation takes advantage of the fact that the Pauli rotations together with the controlled-NOT (CNOT) operation form a complete basis set, which means, every computation can be represented using appropriate Pauli and CNOT operations. This not only shortens but also simplifies the code input, enhancing its portability. An exemplary operation in Pauli representation can be seen in Listing \ref{lst:pauli_rep}, with the Pauli operators of the unitary given as string, characters and index corresponding to the respective Pauli or Identity operation on a specific qubit, a coefficient, and a variable indicating optional parameters. The same circuit in OpenQASM format is also partially displayed and used in the example run depicted in Fig. \ref{fig::ui} \raisebox{.5pt}{\textcircled{\raisebox{-.9pt} {3}}}. Since Qiskit simulators are used in this work for demonstration purposes and many devices accept OpenQASM as IR, the library converting the Pauli representation into OpenQASM is part of the dependencies for the API microservice and ships with the image by default. Users have the option of either entering their code via the live editor or selecting the corresponding file and uploading it. 

\begin{lstlisting}[
float=bp, basicstyle=\footnotesize\ttfamily, caption={Exemplary ansatz Pauli representation. Operators encoded as strings (left), coefficient (middle), and parameter named as number (right).}, captionpos=b, label={lst:pauli_rep}
]
                                        IIXY 1. 1
                                        IXIY 1. 2
                                        IXYI 1. 3
                                        XIIY 1. 4
                                        XIYI 1. 5
                                        XYII 1. 6
                                        IIYX 1. 7
                                        IYIX 1. 8
                                        IYXI 1. 9
                                        YIIX 1. 10
                                        YIXI 1. 11
                                        YXII 1. 12
                                    
\end{lstlisting}

As many circuits performing the algorithm's desired computation need to be parameterized, users must be able to provide the parameters. They can do so either using a dictionary, naming the specific variable to be set and its value, or as a list (array), only providing the variables' values which are then assigned in order of appearance in the circuit. This provision is done on the webpage shown at \raisebox{.5pt}{\textcircled{\raisebox{-.9pt} {4}}}. A prominent example of an algorithm necessitating parameterization is the Variational Quantum Eigensolver (VQE)\cite{vqe_manpreet,vqe, vqe2}. Since parameter optimization is hardware-dependent, a set of optimized parameters obtained on one quantum device cannot be directly fixed into the circuit while ensuring reproducibility across different hardware. However, these parameters can still serve as a good initialization point, reducing the optimization effort on other devices. Therefore, the optimized parameters are included in the result object.

After submission, the provided code is executed via the API on the hardware-specific backend. Following successful execution, the resulting data and metadata are visualized as interactive diagrams or JSON objects as shown in \raisebox{.5pt}{\textcircled{\raisebox{-.9pt} {5}}} of the user interface, with the option of downloading them as CSV files or image files.

\subsection{Q-AIM API}
The Q-AIM API is designed to handle a variety of requests related to both quantum computing tasks and user-specific operations. It is developed using Python and the FastAPI framework and serves as the backbone for processing tasks. Since real quantum hardware is not available for testing, the API utilizes simulators to query as endpoints instead, with the Qiskit library employed for quantum computing task execution using its Qasm Simulator~\cite{QasmSimulator}, a noisy quantum circuit simulator backend.

Primarily, an API comprises public and private endpoints. Public endpoints are accessible without requiring authentication, enabling direct access to the endpoints. Conversely, protected endpoints necessitate authentication via a Json Web Token (JWT), issued by Keycloak, for example. Authentication is facilitated through an authentication function \textit{auth()}, assigned to endpoints requiring authentication as a dependency function using FastAPI's own dependency resolution mechanism. The function issues the query to the identity management using an OAuth2.0 scheme, as described in Section~\ref{sec:authentication_workflow}. For this work, only private endpoints are used to showcase the finely granulated permissions management. These include the endpoint /api/user/me, which retrieves information about the authenticated user. Furthermore, access to endpoints responsible for quantum computing is restricted to authenticated users with appropriate permissions. For illustrative purposes, the prototype offers four more endpoints: for uploading and processing OpenQASM code (\texttt{/api/qc/qasm/\{upload, code\}}), one for each uploading a file and coding on the web page, and the same for code in Pauli representation (\texttt{/api/qc/pauli/\{upload, code\}}). The calculated results are subsequently returned to the Q-AIM frontend as part of the response.

\section{Evaluation}
In the following, we present an evaluation of the integration workflow's key attributes, focusing on its portability and lightweight nature, designed to seamlessly integrate with diverse computing environments. We examine these aspects using different combinations of hardware, software, and hosting paradigms in the following.

\subsection{Test System Setup}
Docker provides a level of abstraction that allows containers to be portable across different environments, including various hardware configurations and operating systems. To assess this portability of the integration workflow's software implementation, our prototype was deployed and tested on three distinct environments: a local machine, an on-premise hosted server, and a cloud instance. These environments span different hardware architectures and operating systems. This multifaceted evaluation aimed to validate Q-AIM's claim of adaptability to diverse computing environments, emphasizing its suitability for individual users with varied system configurations and requirements. The specifications for the different evaluation configurations can be seen in Table \ref{tab:system_settings}.

\begin{table}[h!]
\centering
\caption{Hardware settings for the three evaluation setups.}
\label{tab:system_settings}

\begin{tabularx}{\linewidth}{|l|X|X|X|}
\hline
\textbf{Parameter} & \textbf{Cluster Node} & \textbf{Local Machine} & \textbf{Cloud} \\
\hline
CPU & Intel Xeon E5-2660 v2 & Intel i7-12700H & Intel Xeon E5-2696V4 (vCPU) \\
\hline
Cores & 20 & 20 & 2 \\
\hline
RAM & 128 GB & 32 GB & 8 GB \\
\hline
OS & Rocky Linux 9 & Ubuntu 24.04 & Debian GNU/Linux 12 \\
\hline
Network & Ethernet and InfiniBand (FDR) & Ethernet & Public Internet \\
\hline
\end{tabularx}

\end{table}

First, we demonstrate a proof of concept by deploying on a local machine, i.e., personal computer aimed to simulate real-world scenarios where end-users with diverse machines might seek to utilize the software implementation. The successful deployment on the author's machine confirms that the API functions as intended, providing a sound foundation for further evaluation on more sophisticated hosting paradigm scenarios in the following.

Second, to validate the container's applicability in enterprise settings, we deploy it on real server infrastructure belonging to the Modular Supercomputing and Quantum Computing (MSQC) research group at Goethe University, Frankfurt am Main, Germany. As part of this process, we reconfigured a compute node from the cluster to function as an independent server, ensuring it could operate separately from the main cluster. This emphasizes the applicability in larger research group's and enterprise settings, capable of hosting on-premise solutions, providing full control over the whole workflow.

Third, given the increasing reliance on cloud services in enterprise environments, we also test our solution on Google Cloud using a E2-standard-2 instance, intended for moderate use, providing good trade-off between cost and performance~\cite{GoogleComputePricing}. This deployment is designed to evaluate the feasibility of using the solution in environments with limited computing resources, such as startups, small businesses, or individual developers who often prioritize cost-effective cloud solutions. The successful deployment, despite the limited resources of the cloud instance, underscored the solution's lightweight design and its ability to perform efficiently in resource-constrained cloud environments. Additionally, deploying the solution in the cloud highlights its potential for scalability. Without requiring any modifications to the docker image itself, the container setup can be scaled to more powerful instances, enabling it to handle more demanding workloads as needed.

The consistent behavior observed across different systems and settings underscores the portability and universality of the composed Q-AIM Docker image, substantiating its viability for widespread adoption.

\subsection{Result Discussion}

The ability to deploy and use the sample software implementation on all three distinct infrastructure configurations showcases the portability of the proposed solution. Users are not limited to a single hosting paradigm. From the most straight-forward solution, hosting on personal hardware, to more sophisticated solutions, like cloud-hosting, to ultimately fully on-premise server hosting, every use case can be covered by Q-AIM. 

\begin{figure}[htbp]
\centering
\includegraphics[width=\textwidth]{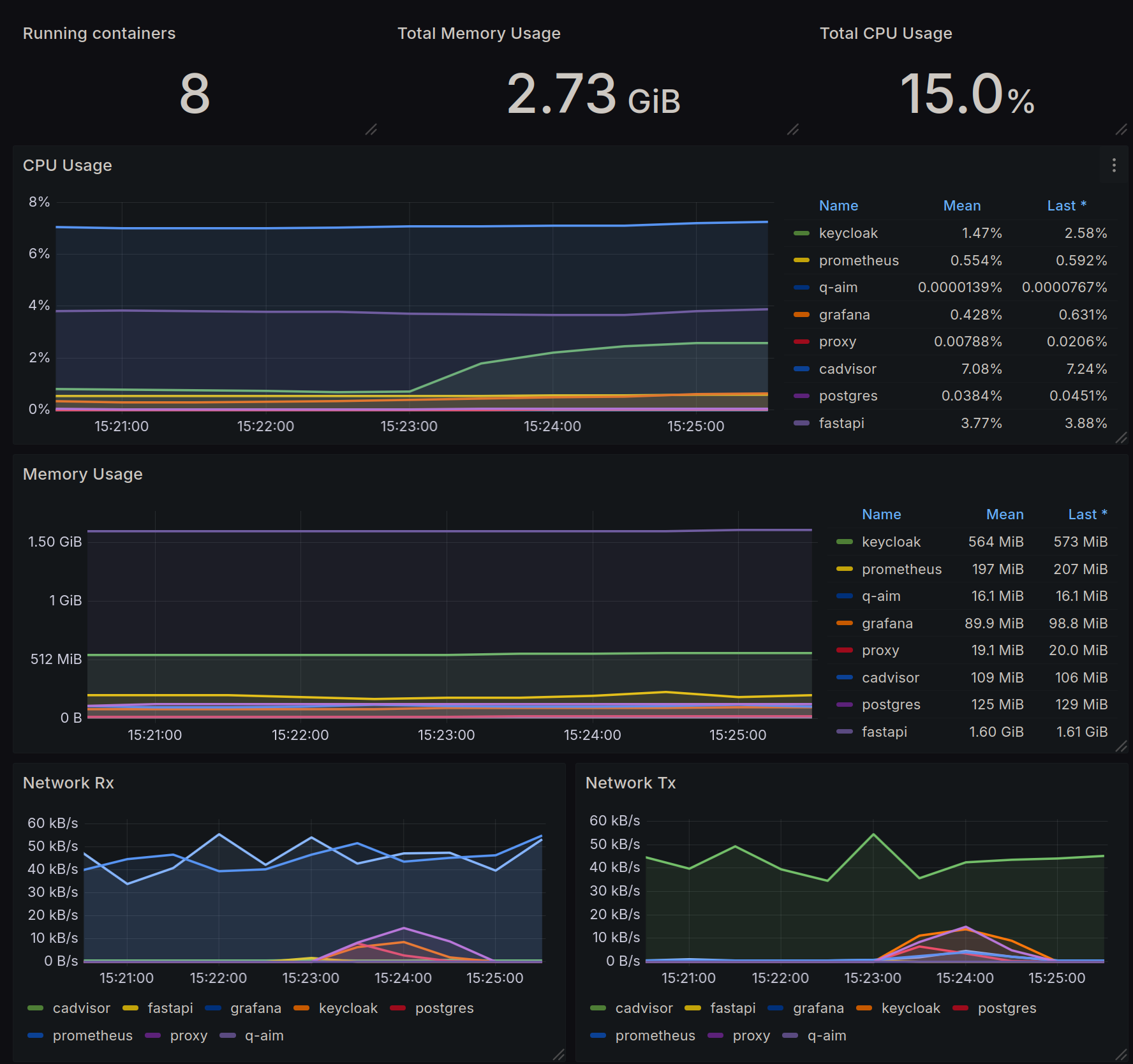}
\caption{Grafana-based Monitoring Dashboard visualizing memory and CPU usage, as well as network traffic (receiving and transmitting) for the different containers in Q-AIM run on the local machine evaluation setup.}
\label{fig::dash}
\end{figure}

Changing the hosting paradigm, e.g. due to higher demand, is just a matter of copying the image and letting it run on the new host, providing the exact same functionality and equal behavior. This reduces the dependency on a particular infrastructure and allows the application of the software to diverse users and use cases. 

The evaluation of the portability made it necessary to deploy the same image on different backends, underlining another key aspect of the docker-based microservice implementation: its reproducibility. The same image of the software, with all its configurations specifically designed for our use case, was easily distributed across multiple infrastructures, which can be understood as providing it to different enterprises. Ultimately, this means enabling other users to use a fully fledged and specifically tailored implementation reduces the overhead of creating a common basis for further research/collaboration.

Another critical aspect of the evaluation pertains to the integration workflow's resource efficiency. 
To investigate resource consumption, the composed Docker container incorporates a resource monitoring software image, cAdvisor\cite{cAdvisorDocker}, as microservice. Running the Q-AIM container automatically starts the monitoring provided by cAdvisor. Utilizing this library, we examined the container's consumption of CPU and memory usage for logging in and running the example from Listing~\ref{lst:pauli_rep} as shown in~Fig. \ref{fig::dash}. Notably, the container exhibited remarkable efficiency, utilizing less than 3 GB of memory in our configuration, whereby Docker uses free memory for caching and frees it as soon as it is needed.

The findings of aforementioned evaluations underscore the integration workflow's software implementation's pivotal attributes, portability across diverse systems and resource-efficient operation. The demonstrated success in real-world scenarios, showed by the seamless deployment on different server infrastructure, positions Q-AIM as a promising solution for users seeking a lightweight, unified, and universally deployable software solution to incorporate quantum computing hardware and offer access to an on-premise device.

\section{Conclusion and Future Work}

We proposed the idea of a single-access software solution integrating quantum resources completely vendor-agnostic. The goal is to enable research groups and small entities willing to procure small quantum devices to streamline the process after procurement until usage and provision. Moreover, the proposed API solution should function as an administration and management tool, providing additional security measures for usage outside of the host's direct scope. Lastly, it must be flexible and portable enough to fit to diverse requirements and infrastructures. Therefore, an open-source containerized microservice solution allows for easy modifications, improved maintainability, and ease of use. Combining all of the above, we introduced Q-AIM, a prototype implementation fitted to our needs of evaluation, providing an interface to a quantum simulator mimicking real integration scenarios to serve as proof-of-concept.

Hosting Q-AIM on diverse infrastructures, ranging from personal machines to cloud instances and up to full on-premise servers, emphasizes its applicability in a wide range of use-cases for different demands. Regardless of the hosting paradigm, Q-AIM ships as self-sustained docker image, incorporating all requirements, demanding no profound knowledge to utilize it as a single access point integration solution for quantum computing devices. Not only the hosting paradigm can be decided upon depending on the host's preferences and demands, but owing to the open-source nature the whole image may be fitted accordingly. This also allows Q-AIM to be fully vendor-independent by enabling specific modifications to be made, tailored for downstream tasks to the connected hardware.

To further showcase our streamlined and portable integration workflow, we will use Q-AIM to integrate the first real quantum computing device for the Modular Supercomputing and Quantum Computing research group at Goethe University in Frankfurt am Main. This milestone will enable efficient management of control and access to the computing service both within and beyond the research group. Future works will include the implementation of error mitigation protocols \cite{em1,em2} within Q-AIM as well as deployment of multi-hybrid quantum algorithms \cite{Jattana2024}.

By facilitating dedicated hardware that bridges the interface to the quantum device’s low-level specifications, Q-AIM enables precise control, allowing users to work at the physical level of quantum hardware—all through a single access point. Moreover, we plan to extend its capabilities to monitor hardware utilization in quantum systems, complementing existing monitoring functions in classical computing. This feature is particularly valuable for scientists engaged in hybrid quantum computing, as it will allow them to seamlessly track the hardware utilization of their quantum-hybrid code. Additionally, by exposing quantum resources via an API, classical computing can leverage these resources as accelerators for specific tasks. To further facilitate hybrid computing, methods such as RPC or customized code constructs like pragmas can be employed to enable asynchronous access to quantum computing resources at runtime.

The production-ready version of Q-AIM aims to serve as a unified access point to quantum hardware, efficiently managing all necessary interactions. This comprehensive approach is designed to streamline integration and control processes, providing scientists and early adopters with a robust and efficient solution for advancing quantum research.

Extensive research is essential in the coming years to enhance devices in the current NISQ era and pave the way for more broadly applicable systems and software that will ultimately surpass it. The development of software like Q-AIM is therefore crucial.

\balance

\bibliographystyle{ieeetran}

\end{document}